\documentclass[final,1p,times]{elsarticle} 
\usepackage{graphicx} 
\usepackage{amssymb} 
\usepackage{amsthm} 
\usepackage{lineno} 
\usepackage{subfigure}

\journal{Nuclear Physics A} 
\begin{document} 

\begin{frontmatter} 


\title{Why is $v_4/(v_2)^2$ larger than predicted by hydrodynamics?\footnote{These proceedings are a condensed version of~\cite{Gombeaud:2009ye}}}


\author[a]{\underline{Cl\'ement Gombeaud}}
\author[a]{Jean-Yves Ollitrault}

\address[a]{Institut de Physique Th\'eorique, CEA/DSM/IPhT,
  CNRS/MPPU/URA2306\\ CEA Saclay, F-91191 Gif-sur-Yvette Cedex.}

\begin{abstract} 

The second and fourth Fourier harmonics of the azimuthal distribution of particles, $v_2$ and $v_4$,
 have been mesured in Au-Au collisions at the Relativistic Heavy Ion Collider (RHIC). The ratio
$v_4/(v_2)^2$ is significantly larger than predicted by hydrodynamics. Effects of partial thermalization 
are estimated on the basis of a transport calculation, and are shown to increase the ratio by a small amount. 
We argue that the large value of $v_4/(v_2)^2$ seen experimentally is mostly due to elliptic flow
 fluctuations. However, the standard model of eccentricity fluctuations 
 is unable to  explain the large magnitude of $v_4/(v_2)^2$ in central collisions.

\end{abstract} 

\end{frontmatter} 

\linenumbers 


\begin{section}{Introduction}
The azimuthal distribution of particles emitted 
in non central nucleus-nucleus collisions at RHIC is a good tool for
understanding the bulk properties of the matter created during the
collisions. 
Near the center of mass rapidity, it can be expanded in the following
Fourier series:
\begin{equation}
\frac{dN}{d\phi} \propto 1+2 v_2 \cos(2\phi)+2 v_4 \cos(4\phi)+\cdots
\end{equation}
where $\phi$ is the azimuthal angle with respect to the direction of the impact parameter.
The large magnitude of elliptic flow, $v_2$, suggests that the matter created in Au-Au collisions 
at RHIC behaves like an almost perfect fluid. However, ideal hydrodynamics predicts $v_4=\frac{1}{2}(v_2)^2$~\cite{Borghini:2005kd}.
 while recent experiments \cite{Abelev:2007qg,Huang:2008vd} 
find $v_4 \simeq (v_2)^2$.
In this talk, I investigate this discrepancy.
\end{section}
\begin{section}{Fluctuations in initial conditions}

\begin{subsection}{Initial eccentricity fluctuations}
Figure\ref{fig:fig1} (left) presents a schematic picture of a 
non central heavy-ion collision (HIC). 
The overlap area between the colliding nuclei has an almond shape,
which generates elliptic flow. This shape is not smooth: 
positions of nucleons within the nucleus fluctuate from one event to
another, even for a fixed impact parameter. 
Therefore, the participant eccentricity, $\epsilon_{PP}$, which is
the eccentricity of the ellipse defined by the positions of
participating nucleons, also fluctuates. 
Since elliptic flow appears to be driven by the participant
eccentricity~\cite{Alver:2008zz}, eccentricity fluctuations translate
into fluctuations of the flow coefficients $v_2$ and $v_4$.  
\begin{figure}[!ht]
\centering
\subfigure{
\includegraphics[width=2.4in]{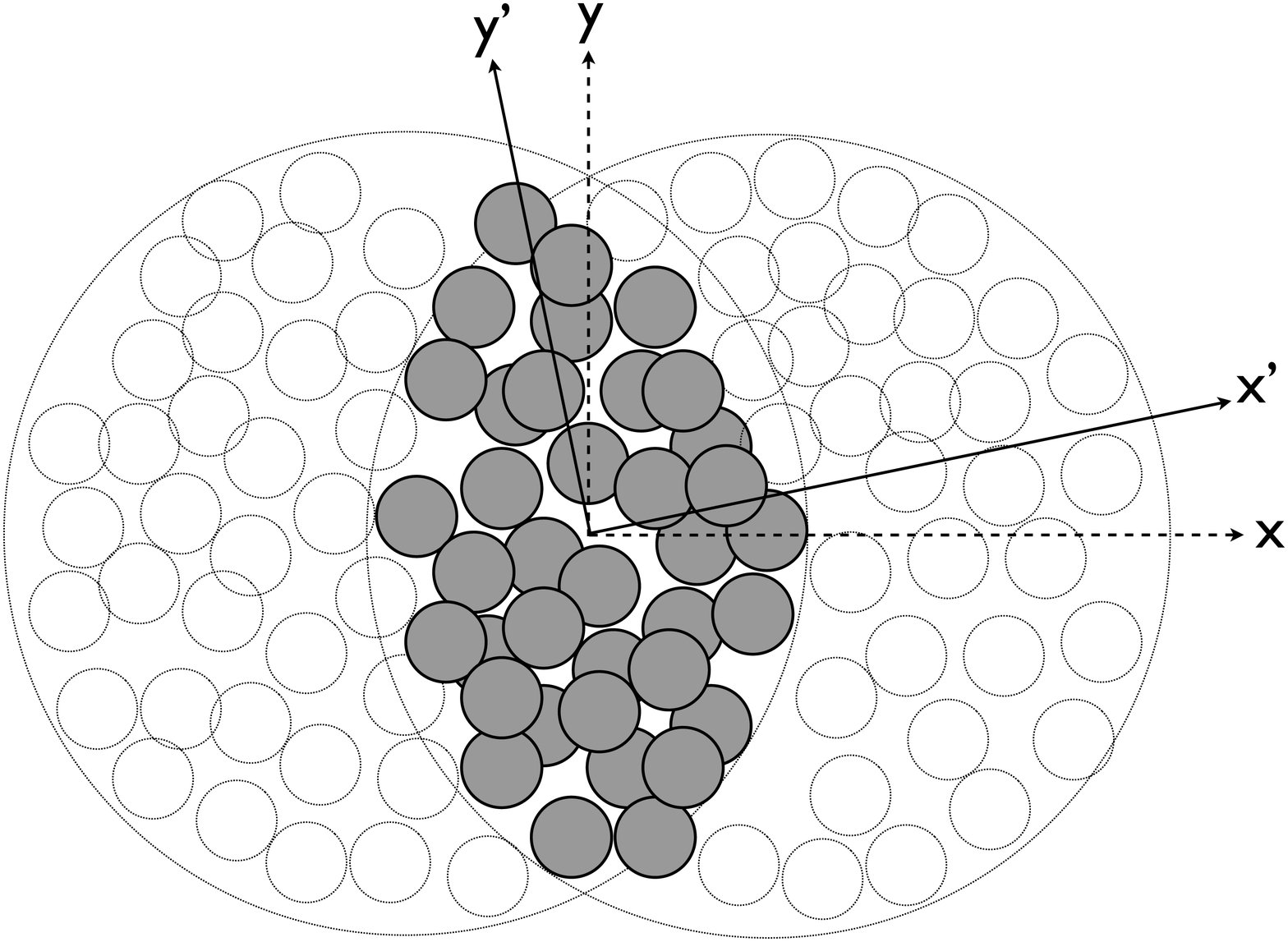}}
\subfigure{
\includegraphics[width=2.7in]{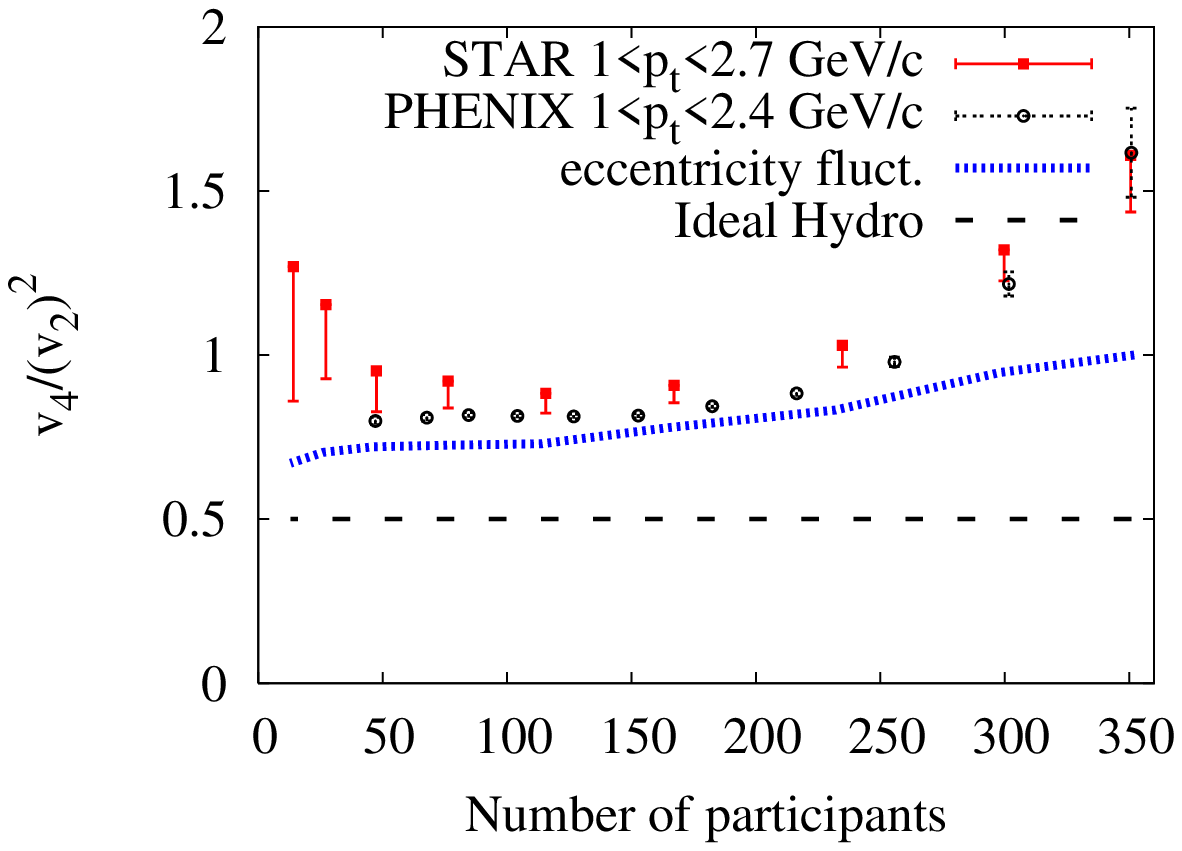}}
\caption{(Color online) Left: Picture of the two frames used for defining the
  initial eccentricity (from~\cite{Alver:2008zz}). The $x$ axis defines the
  reaction plane while the $x'$ axis is the minor axis of the ellipse
  defined by the participants (grey dots). Right: Centrality
  dependence of $v_4/(v_2)^2$: data from 
  STAR~\cite{Y.Bai} and PHENIX~\cite{R.Lacey}; error bars on STAR data
  points are our estimates of nonflow
  errors~\cite{Gombeaud:2009ye}. Lines are  
  predictions from ideal hydro with or without fluctuations.} 
\label{fig:fig1}
\end{figure}
\end{subsection}
\begin{subsection}{Impact of flow fluctuations on $v_4/(v_2)^2$}
There is no direct way of measuring $v_2$ and $v_4$. 
 Analysis methods rely on multiparticle correlations.
Experimentally, $v_2$ can be extracted from the 2-particle correlation and $v_4$ from the 3-particle 
correlation using $\langle \cos(2\phi_1-2\phi_2)\rangle=\langle (v_2)^2 \rangle$ and 
$\langle \cos(4\phi_1-2\phi_2-2\phi_3)\rangle=\langle v_4(v_2)^2
 \rangle$, where angular brackets denote an average value within a
 centrality class. 
Thus any experimental measure of $v_4/(v_2)^2$, obtained using these methods, is rather a measure of 
$\langle v_4(v_2)^2 \rangle/\langle (v_2)^2 \rangle^2$. Inserting the
 prediction from hydrodynamics $v_4=\frac{1}{2}(v_2)^2$, we obtain
\begin{equation}
\left(\frac{v_4}{(v_2)^2}\right)_{\rm measured}=\frac{1}{2} \frac{\langle (v_2)^4 \rangle}{\langle
(v_2)^2 \rangle^2}>\frac{1}{2}.
\end{equation}
We assume that $v_2$ scales like $\epsilon_{PP}$, whose fluctuations
can be estimated using a Monte-Carlo Glauber
model~\cite{Alver:2008zz}. 
The resulting prediction for $v_4/(v_2)^2$ is displayed in 
figure~\ref{fig:fig1} (right). Fluctuations clearly explain most of
the difference between hydro and data. However, experimental data are
still slightly higher than our prediction from fluctuations. We argue
that for peripheral to midcentral collisions, the small residual
difference may be understood in terms of deviations from local
equilibrium. 
\end{subsection}
\end{section}
\begin{section}{Partial thermalization effects}
\begin{figure}[!ht]
\centering
\subfigure{
\includegraphics[width=2.4in]{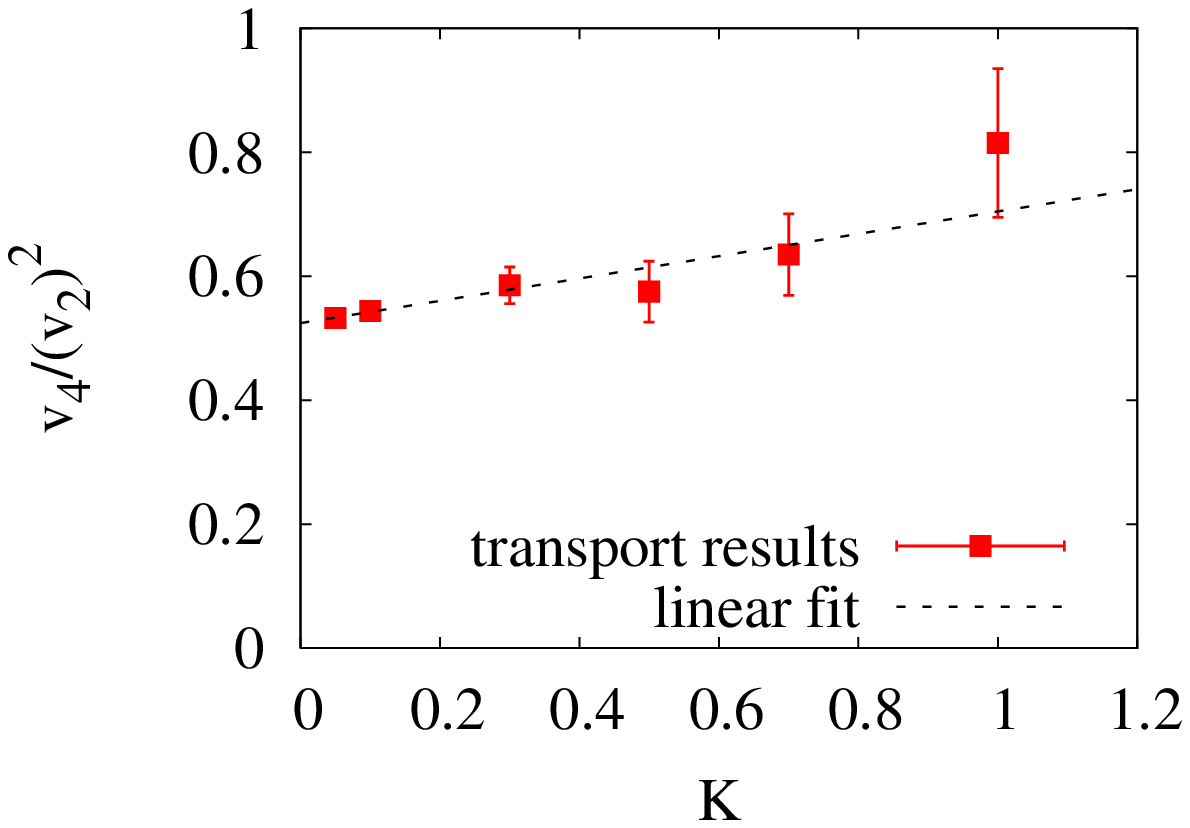}}
\subfigure{
\includegraphics[width=2.7in]{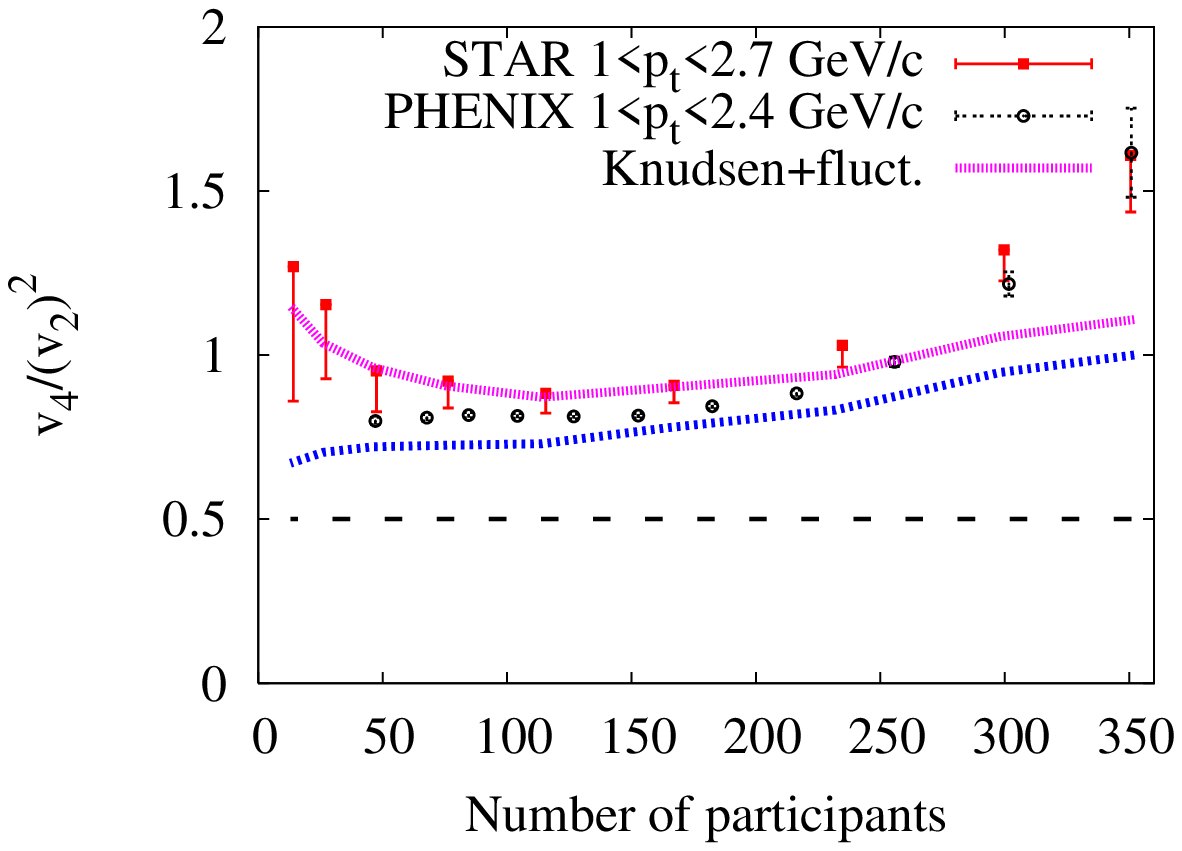}}
\caption{(Color online) Left: Variation of $v_4/(v_2)^2$ with the Knudsen
  number. The point at $K=0$ are obtained using an independant ideal
  hydro calculation.
  Right: Same plot as figure~\ref{fig:fig1} (right), with one
  additional curve showing the effect of the deviation from local
  equilibrium.} 
\label{fig:fig2}
\end{figure}
So far, we have assumed that ideal hydrodynamics correctly describes the expansion of matter created in a HIC. 
But ideal hydro assumes that the system remains in local thermal equilibrium 
(regime where the average number of collisions per particle $n_{coll}$ is large) throughout the evolution. 
In a previous work~\cite{Drescher:2007cd} we have shown that, in order
to reproduce the centrality dependence of elliptic flow,
 the deviation from local thermal equilibrium must be taken into
 account ($n_{coll}\propto 3-5$ would be a typical value
 for Au-Au collisions at top RHIC energy).
 
In the limit of small $n_{coll}$, one expects both $v_2$ and
$v_4$ to scale like $n_{coll}$, so that $v_4/(v_2)^2$ scales like
$1/n_{coll}$: we thus expect 
that the farther the system from equilibrium, the larger
$v_4/(v_2)^2$~\cite{Bhalerao:2005mm}.  In order to have a more
quantitative estimate of this effect, we use a 2+1 dimensional
solution of the relativistic Boltzmann equation to study systems 
with arbitrary $n_{coll}$. We use the Knudsen
number~\cite{Bhalerao:2005mm}, $K\propto 1/n_{coll}$, as a measure of
the  
degree of thermalization of the system. Figure 2 (left) displays the variation of $v_4/(v_2)^2$ with $K$
(see \cite{Gombeaud:2009ye} for details). 
Extrapolation to the hydrodynamic limit $K=0$ yields the value
$0.52$, quite close to the expected $\frac{1}{2}$. For nonzero values of $K$,
$v_4/(v_2)^2$ slightly increases. The effect of this increase on the
centrality dependence is shown in figure 2 (right).
The values of $K$ are borrowed from a previous study  \cite{Drescher:2007cd}.
When both fluctuations and partial thermalization are taken into account,
our calculation slightly overshoots data for midcentral and
peripheral collisions, but the overall agreement is good. We do not yet
understand the large value of $v_4/(v_2)^2$ for central collisions.
\end{section}

\begin{section}{Conclusion}
We conclude that:
1) $v_4$ is mainly induced by $v_2$; 2)
the deviation from local equilibrium has a small effect on $v_4/(v_2)^2$;
3) eccentricity fluctuations explain the observed values of
$v_4/(v_2)^2$, except for the most central collisions which require
further investigation.
\end{section}

\section*{Acknowledgments} 
This work is funded by 'Agence Nationale de la Recherche' under grant
ANR-08-BLAN-0093-01.

\end{document}